\documentstyle[multicol,aps,epsf,pre,graphics]{revtex}

\begin{document}
\title{System-Size Effects on the Collective Dynamics of Cell Populations 
with Global Coupling}
\author{Jun-nosuke Teramae}
\address{Department of Physics, Graduate School of Sciences, Kyoto University,
Kyoto 606-8502, Japan}
\maketitle

\begin{abstract}
 Phase-transitionlike behavior is found to occur in globally coupled 
 systems of
 finite number of elements, and its theoretical explanation is
 provided. The system studied is a population of globally pulse-coupled 
 integrate-and-fire cells subject to small additive noise.
 As the population size is changed, 
 the system shows a phase-transitionlike 
 behavior. That is, there exits a well-defined critical system size above 
 which the system stays in a monostable state with high-frequency activity 
 while below which a new phase characterized by alternation of 
 high- and low frequency activities appears. 
 The mean field motion obeys a stochastic process with state-dependent noise, 
 and the above phenomenon 
 can be interpreted as a noise-induced transition characteristic to such 
 processes.
 Coexistence of high-  and low frequency activities observed in 
 finite size systems
 is reported by N. Cohen, Y. Soen and E. Braun[Physica A{\bf
 249}, 600 (1998)] in the experiments of cultivated heart cells. The present
 report gives the first qualitative interpretation of their
 experimental results.
\end{abstract}



Collective dynamics of coupled dynamical elements represents a central issue of
nonlinear dynamics, and has served as a subject of extensive study
over the last few 
decades. The relevant branches include 
chemical reactions\cite{kuramoto84}, society of living organisms\cite{winfree67,buck81,buck88}, lasers\cite{wang88,wiesenfeld90},
semiconductors\cite{hadley88,wiesenfeld96}, neural
networks\cite{gray89,sompolinsky91,kuramoto92}
and cardiac systems\cite{winfree67,winfree80,glass91}. 
In most theoretical studies, however, the system size is assumed infinite.
While this idealization could be valid for such systems like spatially 
extended chemical reactions, 
there seem to be important practical cases where the finiteness of the
system size should explicitly be taken into account.
Two existing theoretical studies on collective dynamics of populations are 
mentioned here in which
the finiteness of the system size plays a crucial role.
Firstly, Daido\cite{daido90} investigated the collective behavior of 
an inhomogeneous system of 
oscillators focusing on the statistics of fluctuations 
close to the onset of collective motion. 
Secondly, Pikovsky et.al\cite{pikovsky94} studied
coupled bistable elements and showed numerically 
a phase-transitionlike behavior existing only for 
finite systems.
Apart from these theoretical studies,
an interesting
experiment was reported recently by N. Cohen
et.al\cite{cohen98}. They cultivated heart cells in various population sizes
and found that the cells exhibit system-size-dependent behavior.
Some more details of their reports are the following. 
The heart cells extracted from
ventricles of neonatural rats were cultivated, and time series of
spontaneous spiking activities of these cells were recorded. Initially,
the heart cells were made free of mutual contact by a chemical treatment,
but after some time they spontaneously assemble to form subgroups. 
These cell groups 
were cultivated, and the spontaneous
spike activity of the individual cells was observed.
One interesting feature of the experimental results may be stated as follows.
An isolated cell shows random and slow spontaneous spike
activity where the interspike intervals (ISI) distribute sparsely with the
average ISI of $5\sim 10$ seconds. The series of spikes exhibited by the cells 
belonging to relatively small groups are composed of two components
characterized by a distribution with double peaks.
The first low-frequency component is practically the same as that obtained from the spike series for isolated
cells. Another component seems to come from a spike series
which is more regular and its ISI is shorter, i.e., about 1 second.
These low- and high frequency spikes
are visited alternately in time.
If the cell group is sufficiently large, the individual cells exhibit
only high-frequency spike activities of good periodicity. 
The above experimental results are remarkable in that the dynamics of 
the individual cells depends qualitatively on the population size, and still
await theoretical interpretation.
In particular, we would like to know the origin of the transition
which occurs as the population size is changed and whether the mechanism 
involved is universal beyond the particular class of systems of cultivated
heart cells.
The goal of this paper is to provide an answer to these questions. This can be
achieved analytically by using 
a simple dynamical model showing coexistence 
of high- and low frequency activities under suitable conditions. 
Specifically, we employ globally coupled noisy integrated-and-fire model 
which is often
used in the studies of neurodynamics of the brain.
The dynamics of a single cell is given by
\begin{equation}
  \dot x_i(t)=I+\xi_i(t),\quad x_i(t)\leq 1,\quad \left(i=1,\cdots ,N\right),
 \label{eq:integrator}
\end{equation}
where $\xi_i(t)$ is white Gaussian noise with the properties
$\langle\xi_i(t)\rangle=0$ and
$\langle\xi_i(t)\xi_j(t')\rangle=2D\cdot\delta_{ij}\delta(t-t')$, and $N$ is 
the system size which represents the principal control parameter.
At the instant when $x_i$ 
reaches the threshold $x_i=1$, a spiking or firing event is assumed to 
occur for this
cell in such a way that
$x_i$ is immediately reset to a certain value $f$. Thus, 
\begin{equation}
 x_i(t_i^{(n)})=1 \quad \rightarrow \quad x_i(t_i^{(n)}+dt)=f,\quad \left(n=1,\cdots ,N\right),
 \label{eq:reset}
\end{equation}
where $t_i^{(n)}$ is the timing of the spike, the tag $(n)$ indicates the
numbering of the spiking events. 
When a cell fires, this will immediately cause a pulsatile stimulus
on all the other cells in the population.
A given cell will receive a sum of such stimuli coming from various cells.
Assuming that the effect of each stimuli decays exponentially, 
one may conveniently introduce an order parameter $r(t)$ by
\begin{equation}
 r(t)=\frac{1}{N}\sum_{i=1}^{N}\sum_{n;t^{(n)}<t}exp(-\lambda(t-t_i^{(n)})),
 \label{eq:meanfield}
\end{equation}
whose effect is experienced commonly over the cells. 
Since the primary effect of the order parameter should be to lower the
effective threshold for each cell to fire, it would not be unreasonable 
to assume the dependence of the reseting value of $f$ on $r$ like
\begin{equation}
 f(r)=f_0(1-e^{-\beta r}).
 \label{eq:f}
\end{equation}
Note that increasing $r$ implies increasing $f$ with the upper limit 
$f_0$.

Before discussing collective behavior, we show how a single cell behaves
under fixed $r$. 
Equations (1), (2) and (4) determine how the cells behave under a given 
value of $r$.
To characterize statistically the sequence of spikes generated, 
we derive a density distribution function of ISI. Because ISIs
are given by the first passage
time\cite{siegert51,stratonovich63} of the stochastic process 
given by Eq.(\ref{eq:integrator}),
their distribution can be obtained with the standard method
and takes the form 
\begin{equation}
 P_{ISI}(T)=\frac{1-f(r)}{\sqrt{4 \pi DT^3}} \exp(-\frac{(1-f(r)-IT)^2}{4DT}).
 \label{eq:distisi}
\end{equation} 
From this distribution, spike frequency $\omega$ which is defined as the
inverse of mean
ISI, i.e., $1/\langle T \rangle$, becomes
\begin{equation}
 \omega(r)=\frac{I}{1-f(r)}=\frac{I}{(1-f_0)+f_0\cdot e^{-\beta r}}.
 \label{eq:omega}
\end{equation}
One can see that larger/smaller $r$ corresponds to larger/smaller $\omega$
or higher/lower frequency activity. The above results are also consistent
with the experimental facts for the heart cells that when the
cells are isolated the spiking frequency is the lowest and the distribution
of the corresponding ISIs is the broadest.

Two remarks should be given on our model. Firstly, we have taken
into account the effect of mean field on the reseting state by
assuming $f$ to depend on  $r$. However, our choice of the specific form of
$f(r)$ is rather arbitrary. What is important here is 
that the rate of spiking $\omega$ should be an increasing function of
$r$. Under this condition, other choices of $f(r)$ would give qualitatively
the same results.
Secondly, our main goal 
is to understand, analytically if possible, some general features of the 
collective 
dynamics shared commonly by finite-size systems rather than reproducing
precisely the experimental results for the real heart cells, so that
Eqs. (1) to (4), which may not be so realistic for heart cells, would still be
useful enough as a model to work with. 

Equations (\ref{eq:integrator}) to (\ref{eq:f}) are calculated numerically
with some values of $N$, from which the distribution functions for $r$, 
denoted as $P(r,t)$,  
are obtained. Figures 1, 2 and 3 show $P(r,t)$
for the cases of $N=20$, 100 and 200, respectively. 
In order to make clear the difference between the cases of $N=100$
and 200, the data in Figs.2b and 3b are plotted in semilogarithmic scales. 
For sufficiently large system size, the distribution has
a single sharp peak about $r=1.0$ corresponding to coherent 
high-frequency activity. It seems that the distribution shows no qualitative 
change as $N$ becomes even larger. 
In contrast, a remarkable change occurs for smaller $N$. Figures 1
and 2 show the appearance of a new peak at smaller $r$ 
corresponding to low-frequency activity.
The double peaks implies coexistence of the different steady 
states. Actually, the coexistence of high- and low frequency states
is the feature observed experimentally in real heart cells.
Our model gives a fairly well-defined critical population size 
associated with a transition from monostable state to coexistence state. 

In what follows, our analysis proceeds in two steps.
We first derive an evolution equation for the mean field $r(t)$. This can be
achieved by assuming that the mean field $r$ evolves much more slowly than
the evolution of the individual $x_i$ so that the adiabatic approximation
is applicable.
This assumption implies that the random variable $r$ obeys a Markov process
so that the equation for $P(r,t)$ can be written in the form of a
Kramers-Moyal expansion
\cite{kramers40}\cite{moyal49} as,
\begin{equation}
 \frac{\partial P(r,t)}{\partial t}
  =\sum_{n=1}^{\infty}\left(-\frac{\partial}{\partial r}\right)^n D^{(n)}(r,t) P(r,t),
 \label{eq:ck}
\end{equation}
where $D^{(n)}$ is defined as
\begin{equation}
 D^{(n)}(r,t)
  =\frac{1}{n!}\lim_{\tau \rightarrow 0}
  \frac{1}{\tau}\langle [r(t+\tau)-r(t)]^n \rangle |_{r(t)=r}.
 \label{eq:km}
\end{equation}
To find explicit forms of $D^{(n)}$, we calculate $r(t+\tau)-r(t)$ from
Eq.(\ref{eq:meanfield}) for
small $\tau$ under 
the condition that $r(t)=r$. We obtain
\begin{eqnarray}
 r(t+\tau)-r(t) & = & (e^{-\lambda \tau}-1)r+\frac{1}{N}\sum_{i=1}^{N}\sum_{n;t<t^{(n)}<t+\tau}e^{-\lambda(t+\tau-t_i^{(n)})} \nonumber \\
 & = & (e^{-\lambda \tau}-1)r+\frac{1}{N}\sum_{i=1}^{N} n_i(t,t+\tau),
  \label{eq:dr}
\end{eqnarray}
where $n_i(t,t+\tau)$ is the number of spikes of the $i-$th cell during a short
interval $t\sim
t+\tau$. To obtain the last expression in Eq (9),
we use smallness of $\tau$ and replace
$e^{-\lambda(t+\tau-t_i^{(n)})}$ with $1$.
Since $r(t)$ is a slow variable by assumption, no information on the 
dynamics of the 
individual
$x_i$ is relevant except for the timing of their spiking.
The sequence of the spikes is expressed by a Poisson process of the mean
spike rate $\omega(r)$, so that the distribution for $n_i(t,t+\tau)$, denoted by
$P_n(n)$, becomes
\begin{equation}
 P_n(n)=\frac{e^{-\omega(r)\tau}\left( \omega(r) \tau \right)^{n}}{n!}.
 \label{eq:pn}
\end{equation}
Applying Eq.(\ref{eq:pn}) to Eq.(9) and then to Eq.(8), we obtain
\begin{eqnarray}
 D^{(1)} & = & -\lambda r+\lim_{\tau\rightarrow 0}\frac{1}{\tau}\langle n \rangle = -\lambda r+\omega(r), \nonumber \\
 D^{(2)} & = & \frac{1}{2N}\lim_{\tau\rightarrow 0}\frac{1}{\tau}\langle (n-\langle n \rangle)^2\rangle = \frac{\omega(r)}{2N}.
  \label{eq:d1d2}
\end{eqnarray}
Noting that $D^{(n)} \sim O(N^{1-n})$,
and neglecting the terms less than $O(N^{-2})$, Eq.(\ref{eq:d1d2}) 
then reduces 
to a Fokker-Plank equation\cite{Risken89}
\begin{eqnarray}
 \frac{\partial P(r,t)}{\partial t} & = & -\frac{\partial}{\partial r}J(r,t) \nonumber\\
 & = & -\frac{\partial}{\partial r}\left[(-\lambda r+\omega(r))P(r,t)-\frac{\partial}{\partial r}\frac{\omega(r)}{2N}P(r,t)\right].
 \label{eq:fp}
\end{eqnarray}
The above equation is equivalent with the Langevin equation
\begin{equation}
 \dot r(t)=-\lambda r+\omega(r)+\sqrt{\frac{\omega(r)}{N}}\cdot \xi(t),
 \label{eq:langevin}
\end{equation}
where the last term represents noise which is gaussian with the properties
$\langle\xi(t)\rangle=0$ and
$\langle\xi(t)\xi(t')\rangle=\delta(t-t')$. 
From this evolution equation (\ref{eq:langevin}) and the characteristic time scale of
$x_i$, i.e., $T_x=(1-f(r))^2/D$, in which $x_i$ can diffuse
over the interval $[f(r),1]$,  the condition for the adiabatic approximation 
to hold may be expressed as 
\begin{equation}
 \frac{D}{(1-f(r))^2}\gg |\frac{-\lambda r+\omega(r)}{r}|.
\end{equation}
It should be noted
that the noise strength of the stochastic process Eq.(\ref{eq:langevin}) depend on $r$.
Indeed, the fact that the mean field of
a finite size coupled system obeys a stochastic process with state
dependent noise
is a main conclusion of the present paper.

As the second step, we study the behavior of the mean field $r$ 
using Eqs.(\ref{eq:fp}) and
(\ref{eq:langevin}). The systematic part of Eq.(\ref{eq:langevin}), i.e.,
 $-\lambda r+\omega(r)$,
admits a single stable steady state $r=r_c$ as shown in Fig.4.
Thus, for $N=\infty$ the distribution has a single delta
peak at $r=r_c$. The system with smaller $N$ behaves differently,
which comes from the state-dependence of the noise strength.
It is well known that stochastic processes with such noise exhibit
phase-transitionlike behavior called
noise-induced transition\cite{horshemke84}, which means a 
bifurcation exhibited by the locus of the extrema of the distribution 
function. It can be shown that 
Eqs.(\ref{eq:fp}) and
(\ref{eq:langevin}) exhibit such a transition.
The steady distribution for
$r$ is obtained from Eq.(\ref{eq:fp}) where we assume vanishing probability
current, i.e., $J=0$. Thus,
\begin{eqnarray}
 P_{steady}(r) & \propto & \exp (-\Phi(r)) \nonumber \\
 & = & \exp (-\left[2N \int^{r}\frac{-\lambda s+\omega(s)}
 {\omega(s)}ds-\ln \omega(r)\right]),
 \label{eq:steadydist}
\end{eqnarray}
where $\Phi$ represents an effective potential.
Because the extrema of $P$ are identical with the extrema of $\Phi$, 
they can be found from
\begin{equation}
 \frac{d\Phi}{dr}=0
\end{equation}
or
\begin{equation}
N=\frac{\omega'(r)}{2(-\lambda r+\omega(r))}
 \label{eq:extrema}.
\end{equation}
Equation (\ref{eq:extrema}) gives a bifurcation diagram for the extrema
as depicted in Fig.5, where the solid
lines and dotted
line show the loci of maxima and minima of $P_{steady}(r)$, respectively; 
the three
vertical
lines indicate the particular system sizes corresponding to Figs.2, 3, and 4.
The location of the maxima corresponding to each vertical lines thus obtained
analytically is in good agreement with the numerical simulation whose results
are shown in Figs.$1\sim 3$. It is clear from Eq.(\ref{eq:extrema}) and
Fig.5 that there exists a phase transition when $N$ is changed. There
exists a well-defined critical population size $N_c$ for the transition.

In conclusion, the collective dynamics of integrate-and-fire cells
is studied, from which the occurrence of a 
transition at a finite system size 
similar to cultivated heart cells is confirmed. 
In order to understand the origin of the transition,
a stochastic differential equation for the mean field
is first derived.
The transition may be regarded as a noise-induced transition
peculiar to systems with finite size.
Such results 
do not seem confined to a specific model adopted,
but could be observable in 
wide variety of noisy finite-size populations. 

Finally, a few more comments should be given.
It is not intended in the present paper to reproduce 
experimental results (e.g. those by Cohen) quantitatively.
Our main goal was to make clear, with the aid of a relatively simple model,
a certain qualitative feature of the collective
dynamics exhibited by finite-size populations with noise.
In more realistic models, the coupling should be local rather than 
global. In fact, real heart cell interact through electrical coupling
or gap junctions which is local, and this fact is completely ignored 
in the present analysis.
The first important theory on the transitions induced by finiteness
of the system size was developed by Pikovsky et al.\cite{pikovsky94}. 
Some differences between their works and ours are the following. 
Firstly, they assume bistability from the outset for the individual 
elements, while no such assumption is introduced in
our model; bistability appears naturally as a result of collective
dynamics there. Secondly, we succeeded in clarifying analytically
the origin of the transition as a noise-induce transition.

The author thanks Y. Kuramoto for fruitful discussions and critical
reading of the manuscript. He also thanks A. S. Mikhailov for valuable
discussions.




\begin{figure}
 \includegraphics{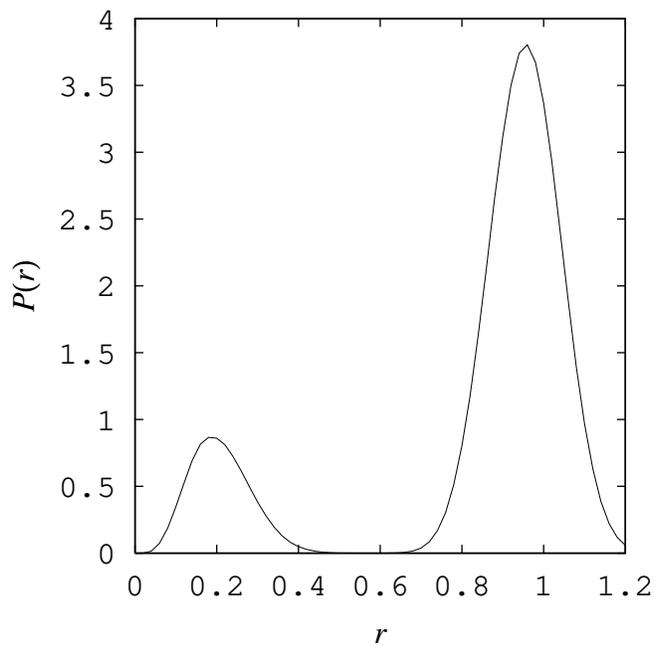}
 \caption{Distribution function for the mean field $r$ calculated numerically 
 from Eq.(\ref{eq:integrator})$\sim$(\ref{eq:f}) with N=20; other
 parameters are $I=0.087, \lambda=1.0, D=10^{-2}, \beta=5.0$ and $f_0=0.92$.}
 \label{fig:20}
\end{figure}

\begin{figure}[]
  \noindent
  {{\large (a)}\includegraphics{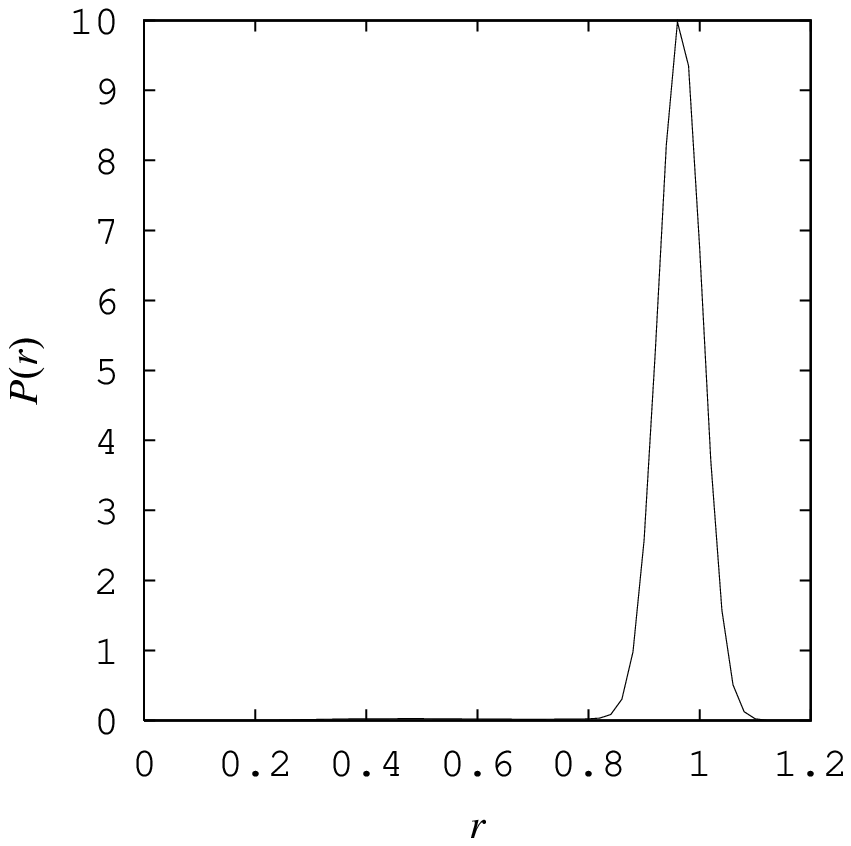}}
  {{\large (b)}\includegraphics{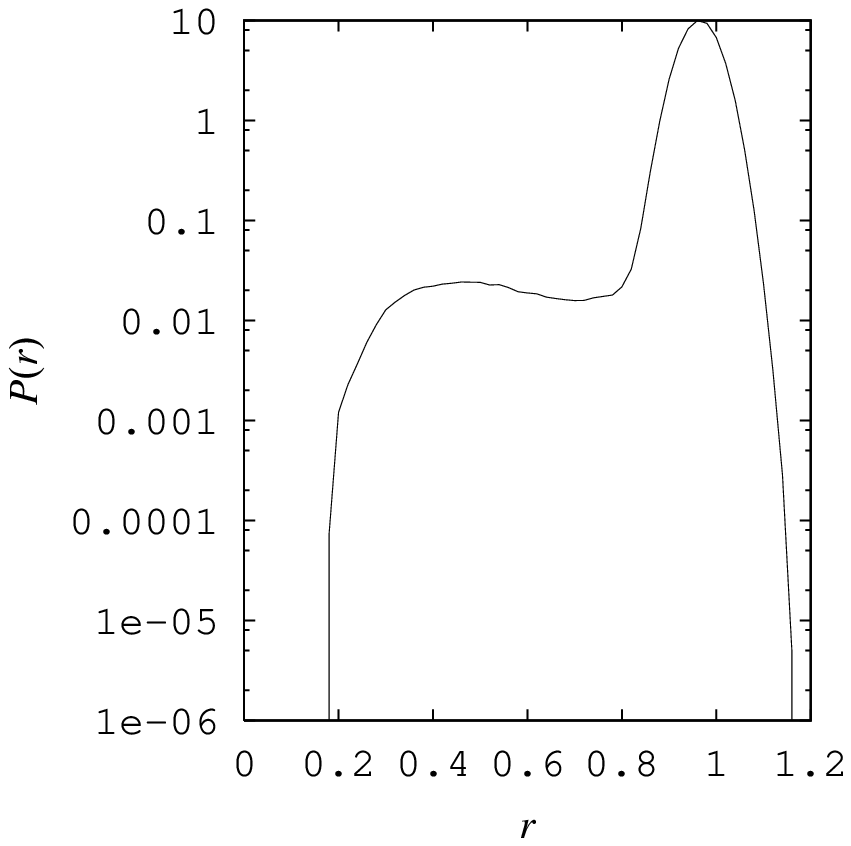}}\\
  \caption{Distribution function for the mean field $r$ calculated numerically 
 from Eq.(\ref{eq:integrator})$\sim$(\ref{eq:f}) with N=100,
 other parameters are same to Fig.1, (b) is plotted with semilogarithmic scale.}
  \label{fig:100}
\end{figure}
 
\begin{figure}[]
  \noindent
  {{\large (a)}\includegraphics{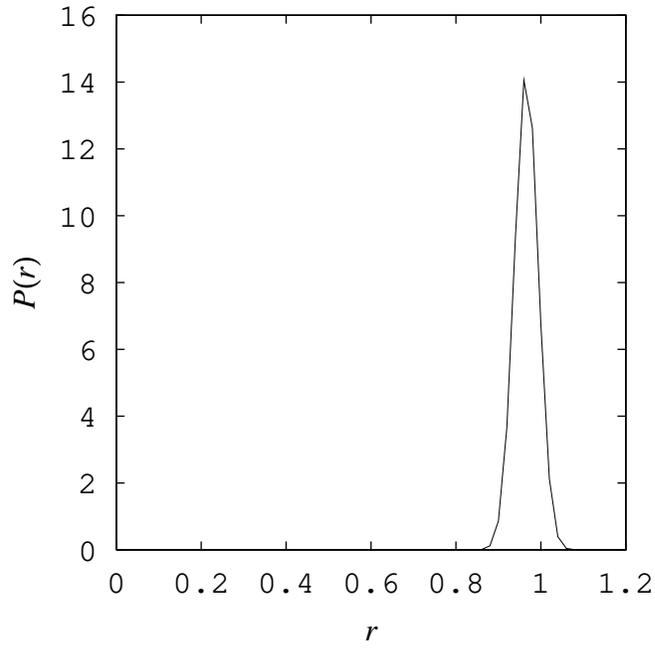}}
  {{\large (b)}\includegraphics{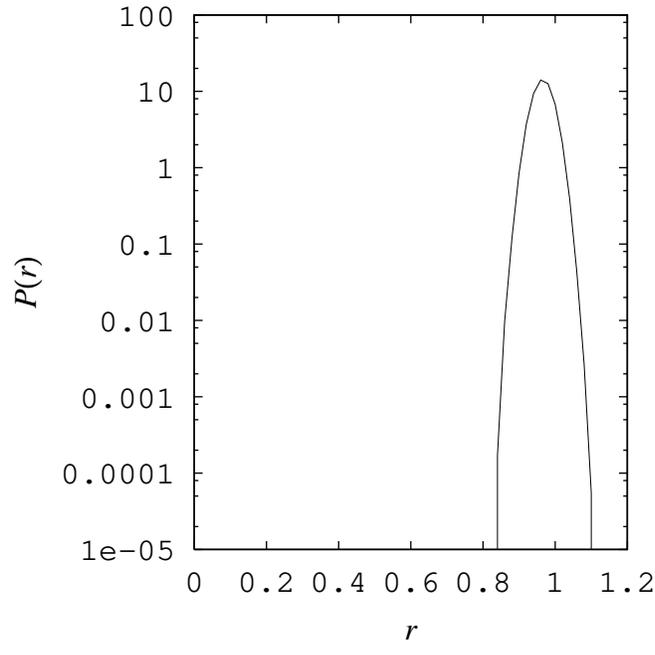}}\\
  \caption{Distribution function for the mean field $r$ calculated numerically 
 from Eq.(\ref{eq:integrator})$\sim$(\ref{eq:f}) with N=200,
 other parameters are same as in Fig.1; in (b) the data are plotted 
 in semilogarithmic scale.}
  \label{fig:200}
\end{figure}
 
\begin{figure}[]
  \includegraphics{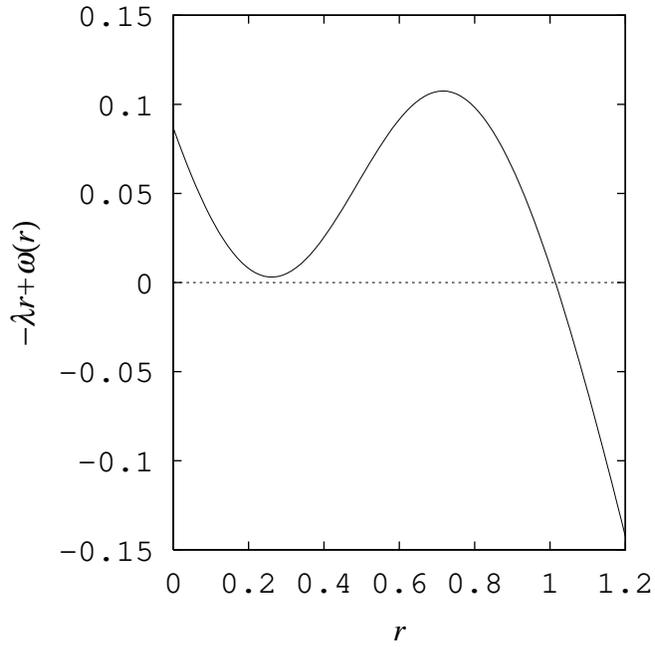}
  \caption{Systematic part of Eq.(\ref{eq:langevin}) as a function of $r$.}
  \label{fig:det_part}
\end{figure}
 
\begin{figure}[]
  \includegraphics{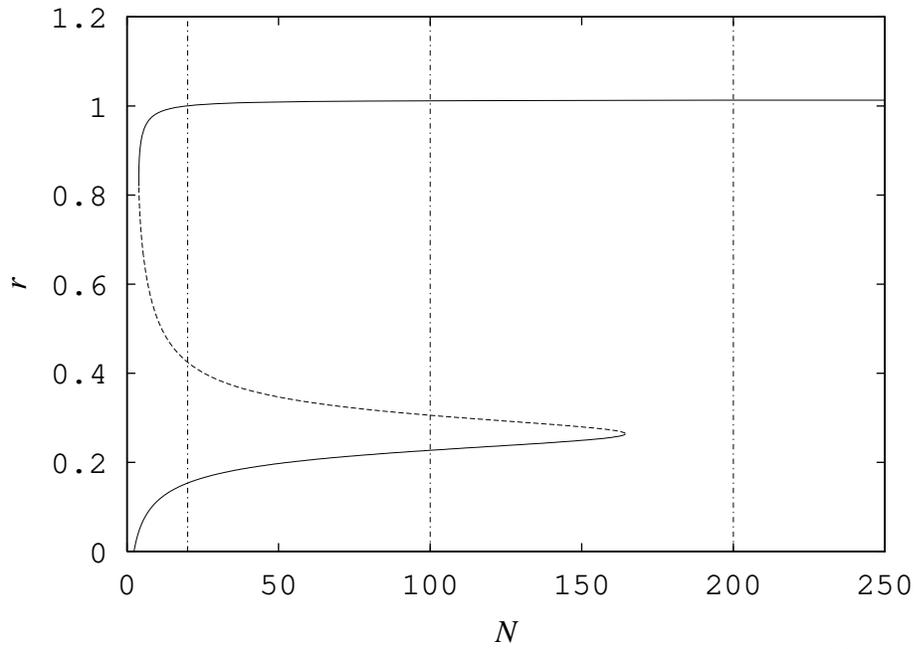}
  \caption{Bifurcation diagram of extrema of the steady distribution
 function $P(r)$ obtained from Eq.(\ref{eq:extrema}). Solid and dotted
 curves indicate maxima and minima, respectively. Vertical lines indicate
 the values of N chosen in Fig.1 $\sim$ 3.}
  \label{fig:bunki}
\end{figure}

\end{document}